\documentclass[apj,twocolumn]{openjournal}
\usepackage{amsmath}
\usepackage{booktabs}

\usepackage[dvipsnames]{xcolor} 
\usepackage[breaklinks,colorlinks,citecolor=blue,urlcolor=blue]{hyperref}

\usepackage{orcidlink}
\usepackage{subfigure}

\renewcommand{\arraystretch}{1.2}  
\usepackage{listings}
\usepackage{color}
\definecolor{dkgreen}{rgb}{0,0.6,0}
\definecolor{gray}{rgb}{0.5,0.5,0.5}
\definecolor{mauve}{rgb}{0.58,0,0.82}
\definecolor{golden}{rgb}{0.86,0.65,0.01}
\defcitealias{Xing2023}{X23}
\defcitealias{Jeena2024}{J24}

\usepackage{soul}
\usepackage{amsmath}
\usepackage{xspace}
\usepackage{graphicx}
\usepackage{enumitem}
\usepackage{amssymb}
\usepackage{xifthen}
\usepackage{hyperref}
\usepackage[normalem]{ulem}
\usepackage{multirow}

\newcommand{\code}[1]{\texttt{#1}\xspace}

\newcommand{\gaia}{\textit{Gaia}\xspace}

\newcommand{\unit}[1]{\ensuremath{\mathrm{\,#1}}\xspace}
\newcommand{\feh}{\unit{[Fe/H]}}
\newcommand{\mfeh}{\unit{[M/H]}}
\newcommand{\vt}{\ensuremath{v_t}\xspace}
\newcommand{\teff}{\ensuremath{T_\mathrm{eff}}\xspace}
\newcommand{\logg}{\ensuremath{\log\,g}\xspace}

\newcommand{\kms}{\unit{km\,s^{-1}}}

\newcommand{\msun}{\unit{M_\odot}}

\newcommand{\thestar}{J\ensuremath{1010+2358}\xspace}

\begin{document}


\author{Pierre N. Thibodeaux\,\orcidlink{0000-0002-3867-3927}$^{1,2}$}
\author{Alexander P. Ji\,\orcidlink{0000-0002-4863-8842}$^{1,2,3}$}
\author{William Cerny\,\orcidlink{0000-0003-1697-7062}$^{4}$}
\author{Evan N. Kirby\orcidlink{0000-0001-6196-5162}$^{5}$}
\author{Joshua D. Simon $^{6}$}

\affiliation{$^1$Department of Astronomy \& Astrophysics, University of Chicago, 5640 S Ellis Avenue, Chicago, IL 60637, USA}
\affiliation{$^2$Kavli Institute for Cosmological Physics, University of Chicago, Chicago, IL 60637, USA}
\affiliation{$^3$Joint Institute for Nuclear Astrophysics}
\affiliation{$^4$Department of Astronomy, Yale University, New Haven, CT 06520, USA}
\affiliation{$^5$Department of Physics, University of Notre Dame, Notre Dame, IN 46556, USA}
\affiliation{$^6$Observatories of the Carnegie Institution for Science, 813 Santa Barbara St., Pasadena, CA 91101, USA}
\email{Corresponding author: pthibodeaux@uchicago.edu}
\title{LAMOST J1010+2358 is Not a Pair-Instability Supernova Relic}

\begin{abstract}
The discovery of a star formed out of pair-instability supernova ejecta would have massive implications for the Population~III star initial mass function and the existence of stars over 100 $M_\odot$, but none have yet been found. 
Recently, the star LAMOST J1010+2358 was claimed to be a star that formed out of gas enriched by a pair-instability supernova.
We present a non-LTE abundance analysis of a new high-resolution Keck/HIRES spectrum of J1010+2358. We determined the carbon and aluminum abundances needed to definitively distinguish between enrichment by a pair-instability and core-collapse supernova. Our new analysis demonstrates that \thestar does not have the unique abundance pattern of a a pair-instability supernova, but was instead enriched by the ejecta of a low mass core-collapse supernova.
Thus, there are still no known stars displaying unambiguous signatures of pair-instability supernovae.
\end{abstract}

\maketitle
\section{Introduction}
\label{sec:intro}

The first Population III stars in the universe are thought to have been extremely massive \citep{Bromm01,Hirano2014}. The top-heavy initial mass function should easily populate the mass range of $140-260\msun$, where stars are expected to explode as pair-instability supernovae (PISNe, e.g., \citealt{Heger2003,Yoon2012,Nomoto2013}).
PISNe are expected to produce a unique abundance pattern with extremely strong odd-even ratios \citep{Heger2002ApJ,Takahashi2018}.
Detecting even one star with a PISN signature would have important implications for the Population~III initial mass function \citep[e.g.,][]{Koutsouridou2024}. 

However, stars exhibiting a clear PISN signature have remained elusive. The first claimed PISN detection \citep{Aoki2014} turned out to have an abundance pattern more consistent with a core-collapse supernova \citep[e.g.,][]{Takahashi2018}.
One difficulty with finding PISNe is that they likely enrich stars to relatively high metallicities $\feh \sim -2.5$ or $-2.0$, so they may be missed in most surveys that primarily target stars with $\feh \lesssim -3$ \citep{Karlsson2008}. The PISN signatures are rapidly erased by any contamination \citep[e.g.,][]{Ji15}, so the signature should be searched for in the presence of contamination by other supernovae \citep{Salvadori2019,Aguado2023}.

Recently, \citet{Xing2023} discovered that the metal-poor star LAMOST J1010$+$2358 (hereafter abbreviated \thestar) is a likely candidate that has preserved a PISN signature.
They found the star has high metallicity ($\feh=-2.4$), low [Mg/Fe], and an extreme odd-even effect including non-detections of Na, Sc, Zn, Sr, and Ba.
The derived abundance pattern is a clear match to a massive (260\msun) PISN\@.
However, \citet[][hereafter \citetalias{Jeena2024}]{Jeena2024} and \citet{Koutsouridou2024} pointed out that the abundance pattern could also be consistent with a core-collapse supernova (CCSN)\@. Key elements needed to strengthen the PISN claim include low abundances of carbon, aluminum, and potassium, which were not measured in the original analysis.

Here we present a new abundance analysis of \thestar using a high signal-to-noise ratio Keck/HIRES spectrum covering an expanded wavelength range. In Section \ref{section:abundance}, we discuss the data and then explain our method of deriving elemental abundances for \thestar. In Section \ref{section:Discussion}, we present the best fitting SN models to our abundances, and we compare our findings to \citet[][hereafter \citetalias{Xing2023}]{Xing2023} in Section \ref{section:comparison}. Contrary to their result, and consistent with the alternate interpretation proposed by \citetalias{Jeena2024}, we find that \thestar is best explained by a CCSN and not a PISN progenitor.

\begin{figure*}
    \centering
    \includegraphics[width=\textwidth]{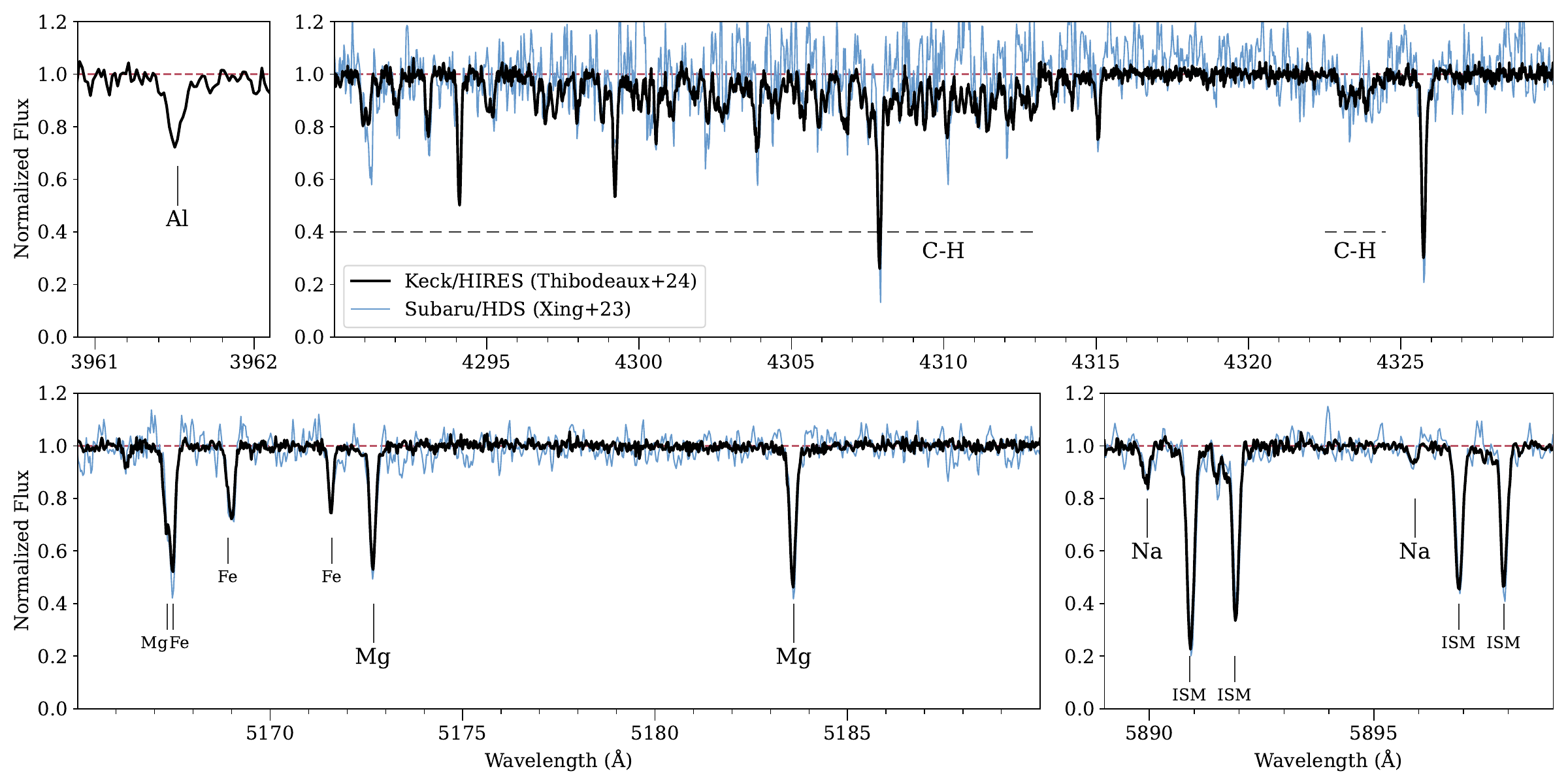}
    \caption{Keck/HIRES spectra (black) of \thestar\ around aluminum, carbon, magnesium, and sodium features. Our re-reduction of \citetalias{Xing2023}'s Subaru/HDS spectrum is plotted in blue for comparison, which shows clear Na D absorption. Our spectra are less noisy, allowing for the measurement of carbon. The HDS spectrum also lacks coverage of the aluminum lines.}
    \label{fig:snapshots}
\end{figure*}

\section{Observations and Abundance Analysis}
\label{section:abundance}
\subsection{Observations}
We obtained a spectrum of \thestar on 2024 January 18 using Keck/HIRES \citep{Vogt_1994} with its red cross-disperser. 
The exposure time was 8100s over 5 exposures with a 1\farcs1 slit and 2x1 binning, resulting in a resolution of $R\sim 36000$.
Data were reduced using \code{MAKEE}~v6.4\footnote{\url{https://sites.astro.caltech.edu/~tb/ipac_staff/tab/makee/}}.
We also observed a nearby telluric standard {(HR~3601)} to verify that the lines used in this analysis are unaffected by telluric absorption.
The signal-to-noise (S/N) is 30 per 0.02~{\AA} pixel at 3950~{\AA} where the blue Al lines are located and 75 per 0.03~{\AA} pixel at 7700~{\AA} where the K 7699~{\AA} line is located.
We used \code{smhr}\footnote{\url{https://github.com/andycasey/smhr/tree/py38-mpl313}} \citep{Casey14} to correct the radial velocity, normalize the echelle orders, and stitch the spectrum.
Figure \ref{fig:snapshots} shows portions of the spectrum around Al\,3961~{\AA}, the Na~D lines, the Mg~b lines, and CH~G band. 
In contrast to \citetalias{Xing2023}, we clearly detect two Na D lines (see discussion in Section \ref{section:comparison}).

\subsection{Stellar Parameters}

We derived stellar parameters using \gaia DR3 photometry and astrometry \citep{GaiaSatellite,GaiaCollaboration2021}. We dereddened the \gaia photometry using the Gaia DR3 relations\footnote{\fontsize{7.5}{7}\url{https://www.cosmos.esa.int/web/gaia/edr3-extinction-law}}, and adopted color-\teff relations from \citet{Mucciarelli2021}. The surface gravity (\logg) was determined using this \teff, applying bolometric corrections from \citet{Casagrande2018}, and the distance from \cite{Bailer-Jones_2018}.  To obtain the microturbulence \vt, we fixed the \logg and \teff and used ATLAS \citep{castelli2004new} with \code{MOOG} in \code{smhr} to balance the [Fe I/H] abundances with respect to the reduced equivalent widths. Our adopted stellar parameters and uncertainties are {compared in Table \ref{tab:paras} to the values from \citetalias{Xing2023}}. 

\begin{table}[]
    \centering
    \caption{Stellar Atmosphere Parameters}
    \begin{tabular}{c c c}
    \hline \hline
         Parameter& \citetalias{Xing2023} & This work \\
         \hline
         $T_\mathrm{eff}$& $5860\pm 120$~K&$5700\pm 200$~K\\
         \logg & $3.6\pm 0.2$ & $4.68\pm 0.3$\\
         \vt & $1.5\pm 0.25$\kms & $1.45\pm 0.2$\kms\\
         \mfeh & $-2.42$ & $-2.5\pm 0.2$
    \end{tabular}
    \begin{center}
        NOTE: The stellar parameters we use for \thestar. Its primary difference from \citetalias{Xing2023} is the \logg value. The model metallicity \mfeh differs from the \feh fit from the spectrum.
    \end{center}
    \label{tab:paras}
\end{table}

To estimate an uncertainty on the parameters, we first looked at \citetalias{Xing2023}, who report $\teff=5860$~K, $\logg=3.6$, $\feh=-2.42$, and $\vt=1.5$~\kms. However, the \citetalias{Xing2023} value for \logg (suggesting that \thestar is a subgiant) is inconsistent with the \logg required by the \gaia DR3 parallax (which implies \thestar is a dwarf). Their reported temperature is consistent with the range of $5700-5900$~K we found based on using different colors (all combinations of $BP$, $G$, $RP$, and $K_s$) of the \gaia photometry. 

We also determined spectroscopic stellar parameters in two ways. First, we followed the method of \citet{Frebel13}, using excitation, ionization, and line strength balance and then applying a correction to the photometric temperature scale to determine $\teff=5540$~K, $\logg=4.20$, $\nu_t=1.62~\kms$.  Second, we {determined stellar parameters including non-LTE (NLTE) effects} using \code{LOTUS} \citep{LOTUS}. {\code{LOTUS} uses pre-calculated curves of growth using the Fe model atom from \citet{Ezzeddine_2016} to determine stellar parameters through excitation, ionization, and line strength balance (see Section \ref{NLTE} for more details)}. We obtained $\teff=5520\pm 165$~K, $\logg=4.8 \pm 0.4$, $\nu_t=0.98 \pm 0.25~\kms$. {Within the 0.3 dex uncertainty, the \code{LOTUS} NLTE \logg is consistent with the \logg from parallax, while the \citet{Frebel13} \logg is consistent when including an extra 0.3 dex offset (see \citealt{Ezzeddine_2020})}.

The effect of stellar parameter uncertainties on the final abundances is dominated by \teff, and we adopt a conservative uncertainty of $\pm 200$~K to cover the range of values that we found. We adopt a \logg uncertainty of 0.3~dex to include our range of \logg values. Our \vt uncertainty is also 0.2~\kms, which is consistent with the \citet{Frebel13} value and includes the upper end of the best-fit \vt range from \code{LOTUS}. 

We adopted $[\alpha/\mathrm{Fe}]=0.0$ and a model metallicity $\mfeh=-2.5$ with an uncertainty of 0.2~dex because the ATLAS model atmosphere grid we used for $[\alpha/\mathrm{Fe}]=0.0$ does not extend below $-2.5$. While this value departs from our final \feh fit, the dependence of the model atmospheres on the \mfeh is weak and does not contribute strongly to our final abundance uncertainty.

\subsection{LTE Abundances}
We conducted an abundance analysis using a similar setup as \citetalias{Xing2023}: \citet{castelli2004new} model atmospheres and \code{MOOG} radiative transfer that assumes local thermodynamic equilibrium (LTE) but includes scattering \citep{Sneden73,Sobeck11}. Linelists for the analysis were adopted from \citet{Ji2020}, with the atomic data originating from \code{linemake} \citep{Placco2021}. We fit equivalent widths and best-fit syntheses in \code{smhr} including formal 5$\sigma$ upper limits for undetected lines (see \citealt{Ji2020} for details).

We also used \code{TSFitPy} \citep{Gerber2023}, which uses \code{Turbospectrum} \citep{Plez2012} and the standard MARCS model atmospheres \citep{Gustafsson2008} to fit selected lines with syntheses. The abundances derived via this method were consistent with the \code{MOOG} abundances for most lines, and we adopt line-by-line differences between \code{MOOG}/ATLAS and \code{Turbospectrum}/MARCS as systematic uncertainties for each line.

We calculated an abundance and an associated error for each detected line in our analysis. The total error $E_{i,\mathrm{tot}}$ for each line $i$ was calculated as the quadrature sum of the statistical error of the \code{MOOG} fits, the abundance difference of the \code{MOOG} and \code{TSFitPy} fits, the abundance difference that comes from increasing one of \teff, \logg, \mfeh, and \vt by the adopted uncertainties, and a minimum systematic error of 0.1 dex for any unmodeled effects. The final LTE abundances are weighted means of the individual line abundances for each species, using inverse variance weights $w_i=1/E_{i,\mathrm{tot}}^2$. The error on the abundance is the uncertainty of the weighted mean, $\sigma_\epsilon=(\sum_i w_i)^{-1/2}$. The uncertainty of the [X/Fe] ratios, $\sigma_\mathrm{XFe}$, differ from $\sigma_\epsilon$ because they account for correlations in X and Fe due to the stellar parameter uncertainties.

\subsection{NLTE Corrections}
\label{NLTE}

{ The assumption of local thermodynamic equilibrium (LTE) fixes the distribution of atoms across their energy levels according to the Saha-Boltzmann equation. However, radiative and collisional interactions can cause the level populations to deviate from LTE. These non-LTE (NLTE) effects can be calculated by solving for statistical equilibrium, which influences the abundances inferred from the spectrum.}

{ We used \code{TSFitPy}  to determine NLTE corrections for individual lines \citep{Gerber2023}. \code{TSFitPy} uses precomputed NLTE departure coefficient grids calculated from the following model atoms:  Na \citep{Larsen2022}, Mg \citep{Bergemann2017}, Ca \citep{Mashonkina2017,Semenova2020}, Ti \citep{Bergemann2011}, Mn \citep{Bergemann2019}, Fe \citep{Bergemann2012a,Semenova2020}, Co \citep{Bergemann2010a,Yakovleva2020}, Ni \citep{Bergemann2021,Voronov2022}, Sr \citep{Bergemann2012c}. } The NLTE correction for a single line is computed by taking the difference of the NLTE abundance and the LTE abundance from {fitting the spectrum with}  \code{TSFitPy}. The NLTE corrections for the Aluminum lines were calculated seperately from \citet{Nordlander17}. The total NLTE correction uses the \code{MOOG} line weights.

Our final abundances for \thestar are the weighted average \code{MOOG} abundances with the weighted average \code{TSFitPy} NLTE correction. These final abundances are listed in Table \ref{tab:my_label} as $\log \epsilon$, $[\mathrm{X/H}]$, and $[\mathrm{X/Fe}]$ alongside their error and the NLTE corrections (which are already incorporated into the abundances). We adopt solar abundances from \citet{EMagg2022}, which are revised from \citet{Asplund09}.

\begin{table}[]
    \centering
    \caption{Chemical Abundances}
    \begin{tabular}{c c c c c c c c}
    \hline \hline
        ID & N & log $\epsilon$ & [X/H] & [X/Fe] & $\sigma_{\epsilon}$ & $\sigma_\mathrm{XFe}$ & $\Delta_\mathrm{NLTE}$\\
        \hline 
		CH&2&$6.10$&$-2.46$&$+0.16$&$0.26$&0.21&$\dots$\\
		O I&1&${<}{8.59}$&${<}{-0.18}$&${<}{+2.44}$&limit&limit&$\dots$\\
		Na I&2&$2.25$&$-4.04$&$-1.42$&$0.13$&0.08&$-0.03$\\
		Mg I&4&$4.30$&$-3.25$&$-0.63$&$0.11$&0.08&$+0.02$\\
		Al I&2&$2.70$&$-3.73$&$-1.11$&$0.20$&0.15&$+0.61$\\
		K I&1&${<}{2.93}$&${<}{-2.21}$&${<}{+0.41}$&limit&limit&$\dots$\\
		Ca I&4&$3.33$&$-3.04$&$-0.42$&$0.08$&0.08&$+0.04$\\
		Sc II&2&$-0.18$&$-3.25$&$-0.63$&$0.19$&0.21&$\dots$\\
		Ti I&3&$2.28$&$-2.66$&$-0.04$&$0.15$&0.10&$+0.20$\\
		Ti II&5&$2.05$&$-2.89$&$-0.27$&$0.07$&0.12&$+0.15$\\
		V I&1&${<}{1.52}$&${<}{-2.37}$&${<}{+0.25}$&limit&limit&$\dots$\\
		V II&1&${<}{2.08}$&${<}{-1.81}$&${<}{+0.81}$&limit&limit&$\dots$\\
		Cr I&2&$2.73$&$-3.01$&$-0.39$&$0.15$&0.08&$\dots$\\
		Mn I&2&$2.52$&$-3.00$&$-0.37$&$0.13$&0.11&$+0.23$\\
		Fe I&73&$4.88$&$-2.62$&$0.00$&$0.02$&0.02&$+0.09$\\
		Fe II&6&$4.78$&$-2.72$&$-0.10$&$0.07$&0.16&$+0.01$\\
		Co I&2&$2.18$&$-2.77$&$-0.14$&$0.15$&0.09&$+0.16$\\
		Ni I&1&$3.57$&$-2.67$&$-0.05$&$0.20$&0.12&$+0.36$\\
		Zn I&1&${<}{2.64}$&${<}{-1.92}$&${<}{+0.70}$&limit&limit&$\dots$\\
		Sr II&2&$-1.19$&$-4.06$&$-1.44$&$0.13$&0.16&$+0.07$\\
		Ba II&1&${<}{-1.52}$&${<}{-3.70}$&${<}{-1.08}$&limit&limit&$\dots$\\
		Eu II&1&${<}{-0.85}$&${<}{-1.37}$&${<}{+1.25}$&limit&limit&$\dots$\\
        \hline
    \end{tabular}
    \begin{center}
        NOTE: The \citet{EMagg2022} solar abundances were used for normalization. NLTE corrections are already applied to the $\log\epsilon$ and other abundance values. [X/Fe] has an additional NLTE correction for [Fe/H] already applied. We use the species Fe\,I, Ti\,II, and V\,II for SN model fitting.
    \end{center}
    \label{tab:my_label}
\end{table}

\begin{figure*}
    \centering
    \includegraphics[width=\textwidth]{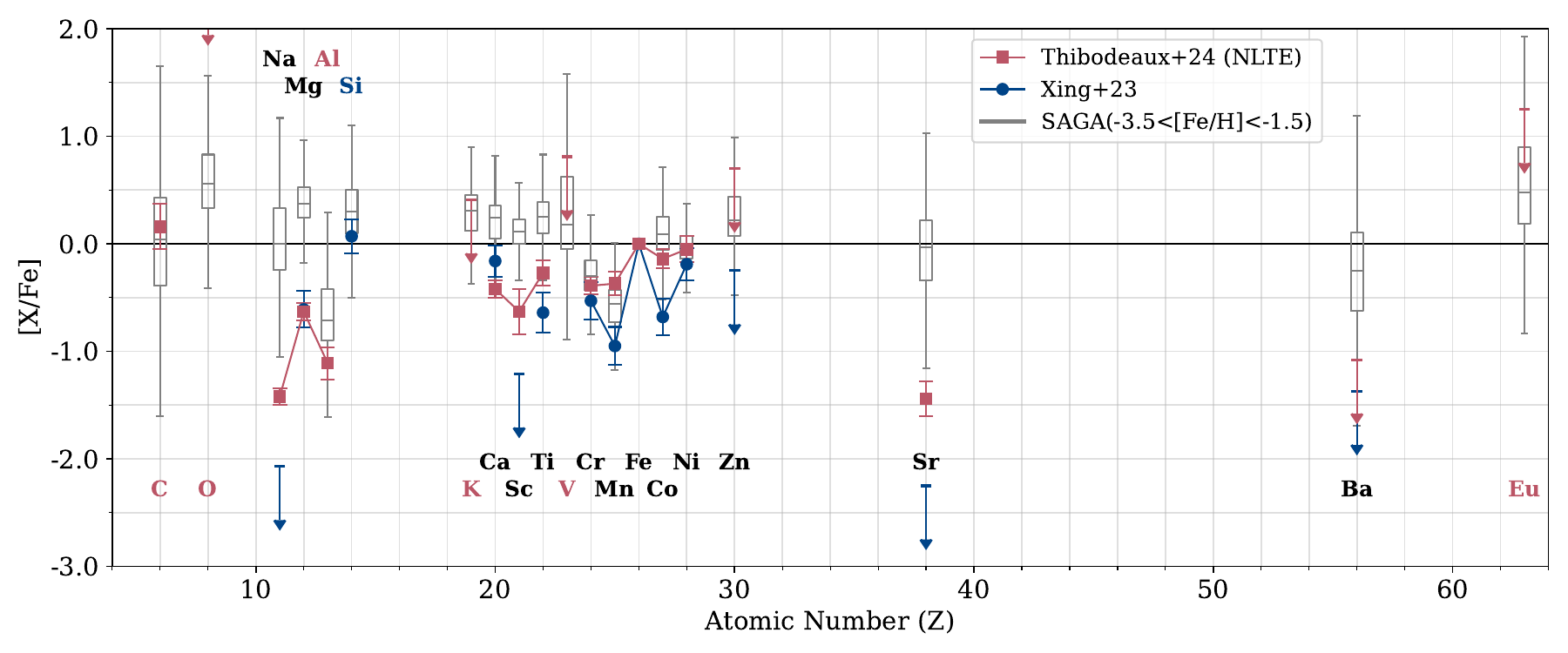}
    \caption{[X/Fe] values for \thestar against those from \citetalias{Xing2023} and against SAGA abundance ranges. The uncertainties on our abundances are $\sigma_\mathrm{X/Fe}$ from Table \ref{tab:my_label}. The SAGA solar normalizations have been adjusted from \citet{Asplund09} to \citet{EMagg2022}. We have applied our NLTE correction to the SAGA aluminum value. The colors of the element labels correspond to the elements only measured by us (red), only measured in \citetalias{Xing2023} (blue), or in both (black).}
    \label{figure:saga}
\end{figure*}

The final abundances are plotted in Figure \ref{figure:saga} as $[\mathrm{X/Fe}]$ in comparison to the original \citetalias{Xing2023} abundances and the distribution of abundances for stars with $-3 < \mbox{[Fe/H]} < -1.5$ from the SAGA database \citep{Suda2008}. We adjust the SAGA aluminum abundances by our adopted NLTE correction since they were mainly computed in LTE\@. Compared to the SAGA stars, \thestar is low in magnesium, calcium, and barium. Additionally, we find that the star is low in sodium, scandium, titanium, and strontium, though not as low as \citetalias{Xing2023} reported.

\section{SN Yield Fits}
\label{section:Discussion}

We compare our abundances to a grid of 16800 CCSN nucleosynthesis models \citep{Heger10} and 14 PISN models \citep{Heger2002ApJ}. We determine the best-fit model by minimizing the mean absolute deviation between the abundances of \thestar and those of the models for elements $Z=6-30$, allowing our [X/H] measurements to scale with a dilution parameter. Compared to a chi-squared ($\chi^2$) minimization, this penalizes individual bad element fits less severely, though we still use the $\chi^2$ to qualitatively visualize the range of well-fitting models. We reject any solutions that violate an upper limit, though this does not affect our final best fit.

\begin{figure*}
    \centering
    \subfigure[Our SN fitting of the abundances from \citetalias{Xing2023}. Their result of a $260 M_\odot$ PISN is recovered as the best solution, though CCSN models are possible within $5\sigma$ of their measurements, as first noted by \citetalias{Jeena2024}.]{\includegraphics[width=\textwidth]{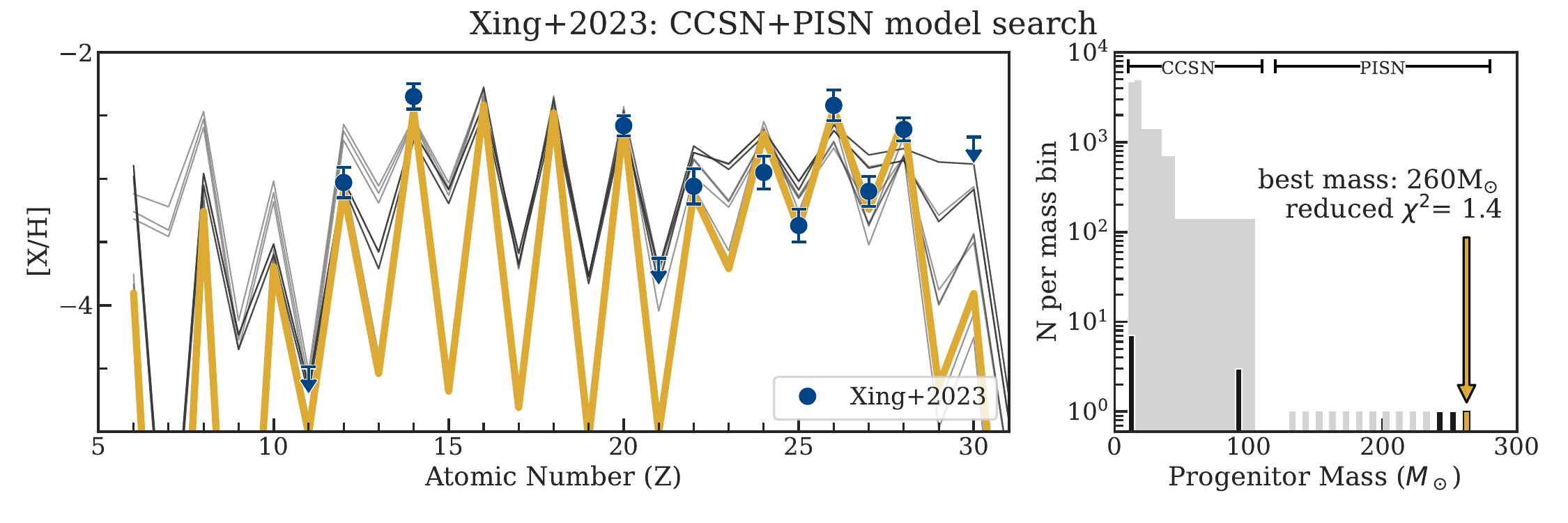} \label{fig:CCSN_bestfitx}}
    \subfigure[The best-fitting model over all surveyed models is an 11 $M_\odot$ CCSN model \citep{Heger10}. The reduced chi-squared of this fit is 2.8, and the reduced chi-squareds of the other "well-fitting" models range between 2.2 and 6.4. PISNe are completely excluded.]{\includegraphics[width=\textwidth]{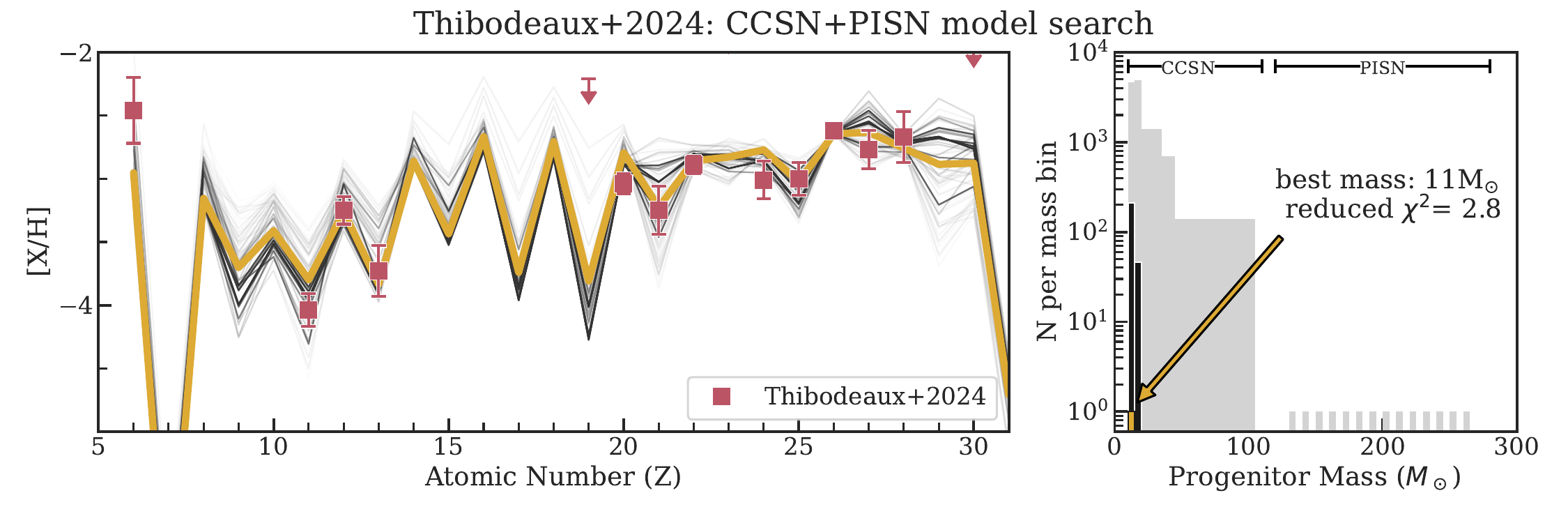} \label{fig:CCSN_bestfita}}
    \subfigure[The best-fitting PISN model \citep{Heger2002ApJ} compared to our abundances. The reduced $\chi^2$ of this fit is 28.8, which is clearly a worse fit than the best CCSN model, especially for carbon (Z=6), sodium (Z=11), aluminum (Z=13), calcium (Z=20), scandium (Z=21), chromium (Z=24), and cobalt (Z=27).]{\includegraphics[width=\textwidth]{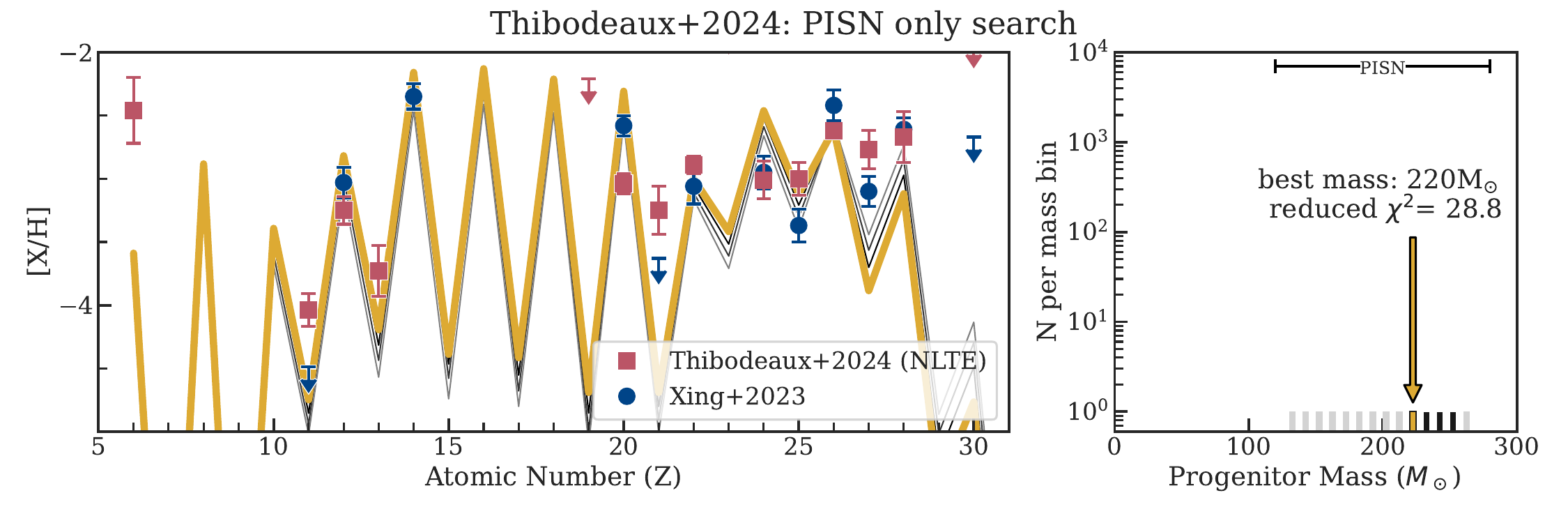}\label{fig:CCSN_bestfitb}}
    \caption{Our model search. For the left plot in each subfigure, the best model (gold) is the one that minimizes the mean absolute error from the [X/H] abundances. This includes minimizing over a dilution parameter, which would shift the [X/H] of the model up or down. The uncertainties in these plots correspond to $\sigma_{\epsilon}$ in Table \ref{tab:my_label}. The black ``well-fitting" models are those whose $\chi^2$ fall within a $5\sigma$ range of the minimum $\chi^2$ value, and their opacities correspond to their strength-of-fit. Red squares correspond to the abundances found in this work, and the blue circles correspond to the abundances of \citetalias{Xing2023}. On the right of each subfigure, the progenitor masses of all the models are plotted as the grey histogram. The progenitor masses of the well-fitting models are plotted in black, and the best-fitting model is plotted in yellow.}
    
\end{figure*}

We first verified that our SN fitting program worked by using the original \citetalias{Xing2023} abundances to find the well-fitting (within $5\sigma$) models from our grid, as shown in Figure \ref{fig:CCSN_bestfitx}. In this plot, we display models with a $\chi^2$ value within $5\sigma$ of the optimal value. The best-fitting model is displayed in gold, whereas the well-fitting models are plotted in black, both in the Z vs.\ [X/H] plot and in the progenitor mass histogram. The total search space of progenitor masses is also plotted in gray in the right plot. The reduced $\chi^2$ is calculated as $\chi^2/(\mathrm{d.o.f})$, where the degrees of freedom is the number of elements fit (excluding upper limits) minus the number of model parameters (4 for CCSN, 2 for PISN).

With the \citetalias{Xing2023} abundances, we recover their solution of a 260~$M_\odot$ PISN model as the best-fitting solution but also found some CCSN models among the well-fitting models. These models are not within $4\sigma$ of the optimal $\chi^2$ fit, so they are not preferred; however, it shows that a CCSN model could be found to explain their data. This possibility was first pointed out in \citetalias{Jeena2024}, as was the fact that \thestar's low $[\mathrm{Mg/Fe}]$ does not preclude it from being enriched by a CCSN. 

With our updated abundances, we find the best-fitting model (reduced $\chi^2=2.8$) to be an 11 $M_\odot$ CCSN, as shown in Figure \ref{fig:CCSN_bestfita}.
As suggested by \citetalias{Jeena2024}, the measurements of aluminum (Z=13) and carbon (Z=6) clearly distinguish between a CCSN and PISN\@. However, the rest of our abundances are consistent with a CCSN model as well, and we recover the same result when we remove carbon and aluminum from our fits. Even when excluding C and Al, PISN models are completely excluded from the well-fitting models. 

When we restrict our search to just PISN models, as displayed in Figure \ref{fig:CCSN_bestfitb}, the best-fitting model is a 220 $M_\odot$ PISN. However, this is a lower-quality fit (reduced $\chi^2 = 28.8$). With the exception of sodium (Z=11) and magnesium (Z=12), we do not see a strong odd-even pattern in our measured abundances. The iron peak elements also lack the strong odd-even pattern that we would expect for a PISN\@. The difference in conclusions between our analysis and \citetalias{Xing2023} stems from disagreements in the abundance measurements, most notably sodium, calcium (Z=20), scandium (Z=21), titanium (Z=25), and cobalt (Z=27).

{
Following \citetalias{Xing2023}, our conclusions are based on the PISN and CCSN model grids from \citet{Heger2002ApJ} and \citet{Heger10}.
We note that we explored a much wider range of 1D metal-poor CCSN and PISN yield models described in \citet{Ji2024} (\citealt{Limongi2012,Limongi2018,Nomoto2013,Grimmett2018MNRAS,Ritter2018,Ebinger2020ApJ}), finding identical conclusions.
Most of these models assume spherical symmetry, zero metallicity, and no rotation. Such models are appropriate for PISNe (though see \citealt{Yoon2012} for discussion of rotation), while the CCSN models are heavily parameterized with different explosion energies and mixing/fallback prescriptions.
The impact of assuming 1D is harder to assess. While significant progress has been made in recent years on 3D CCSN simulations \citep[e.g.,][]{Sieverding_2020,Sieverding2023,Wang2023}, they have all been run on solar metallicity progenitors.
This said, existing 3D models suggest that nucleosynthesis effects are restricted to elements heavier than Ca ($Z \geq 20$) \citep{Sieverding_2020}. Even if we only consider elements with $Z < 20$ and Fe, we would still conclude that PISNe are excluded while CCSNe are allowed.}

\section{Comparison to {the Literature} }
\label{section:comparison}

Because it substantially affects the interpretation, we investigated possible reasons for our abundance differences compared to  \citetalias{Xing2023}. First, they adopted different stellar parameters; most importantly, their \logg of $3.6$, differs from our value of $4.68$. Though our spectroscopic and photometric parameters disagree with this result, we analyzed our spectrum with their stellar parameters and in LTE\@. The discrepancies between our values of scandium and magnesium are attributable to the different stellar parameters, but other elements remain inconsistent. 

Our uncorrected, LTE abundances have a slightly stronger odd-even effect in the Fe peak elements; however, repeating our SN fitting procedure yielded the same result, with the same exclusion of PISNe from the well-fitting models. We could find no changes in the analysis of our spectrum that would reproduce the low Na or Si detection reported by \citetalias{Xing2023}.

In particular, our Na D detection is in direct conflict with the clear non-detection shown in their extended data figure 1.
We thus downloaded the data used in \citetalias{Xing2023} from the SMOKA archive \citep{Baba2002} and reduced it in \code{IRAF} \citep{IRAF1,IRAF2,IRAF3} with the HDS routines. The S/N of the HDS spectrum is lower than the HIRES spectrum, but we clearly see the two stellar Na~D lines, as well as the stronger interstellar absorption components (Fig~\ref{fig:snapshots}). The stellar Na lines are also visible on the raw 2D science image, where we identify some cosmic rays near the Na D lines. We suspect cosmic ray removal could have impacted the data reduction in \citetalias{Xing2023}. Furthermore, the 2D inspection shows that the star is not centered on the slit, making sky subtraction more difficult. As a final check, we measured equivalent widths of the lines we analyzed in our reduction of the HDS spectrum, finding no significant differences after accounting for spectrum noise.
In any case, our higher S/N Keck/HIRES spectrum should provide a more reliable abundance analysis.

{Recently, \citet{skuladottir_2024} published an independent analysis of a VLT/UVES spectrum of \thestar. Their abundance measurements agree with our analysis, and they additionally consider combinations of multiple supernova progenitors. Their work favors a combination of Population II and III CCSNe to explain the abundance pattern of \thestar.}

\section{Conclusion}
\label{section:conclusion}
We present a new abundance analysis of the PISN candidate, \thestar. We collected a Keck/HIRES spectrum and calculated abundances using both equivalent width fitting and spectral synthesis fitting. Our new abundances confirm that the abundance pattern of \thestar originated in a core-collapse supernova, and not a pair-instability supernova, as previously reported. Consistent with the suggestion of \citet{Jeena2024}, our new carbon and aluminum measurements strongly favor a CCSN explanation over a PISN explanation. We find other discrepancies in our abundances compared with those determined by \citet{Xing2023}. Our result means that a true PISN candidate is yet to be found.


\section*{acknowledgments}

P.T. and A.P.J. acknowledge support by the National Science Foundation under grants AST-2206264 and AST-2307599. W.C acknowledges support from a Gruber Science Fellowship at Yale University. E.N.K.\ acknowledges support from NSF CAREER grant AST-2233781.
This research has made use of the Keck Observatory Archive (KOA), which is operated by the W. M. Keck Observatory and the NASA Exoplanet Science Institute (NExScI), under contract with the National Aeronautics and Space Administration.
Based in part on data collected at Subaru Telescope and obtained from the SMOKA, which is operated by the Astronomy Data Center, National Astronomical Observatory of Japan.

The authors wish to recognize and acknowledge the very significant cultural role and reverence that the summit of Maunakea has always had within the Native Hawaiian community. We are most fortunate to have the opportunity to conduct observations from this mountain.



\bibliography{main}{}
\bibliographystyle{aasjournal}

\end{document}